\documentclass[a4paper]{article}
\usepackage[dvips]{graphicx}

\begin{document}
\begin{center}
\begin {title}
\title{\bf{CHANGE OF PRIMARY COSMIC RADIATION NUCLEAR
COMPOSITION IN THE ENERGY RANGE  $10^{15}  -  10^{17}$  eV AS A
RESULT OF THE INTERACTION WITH THE INTERSTELLAR\\ COLD BACKGROUND
OF LIGHT PARTICLES}}
\end {title}
\end{center}
\begin{author}
\author{  T.T.~Barnaveli\hspace*{2mm}(*), ~T.T.~Barnaveli (jr) ,~
N.A.~Eristavi , ~I.V.~Khaldeeva }\\

 {\small\it {Institute of Physics, Georgian Acad.Sci.,
 Tamarashvili 6, Tbilisi 380077, Georgia}}\\

(*) e-mail: mantex@caucasus.net  ;
            barnaveli@hotmail.com \\
\end{author}


      In this paper the updated arguments in favor of a simple
model, explaining from the united positions all peculiarities of
the Extensive Air Shower (EAS) hadron $E_h (E_0)$ (and muon
$E_{\mu}(E_0)$) component energy fluxes dependence on the primary
particle energy $E_0$ in the primary energy region $10^{15} -
10^{17}$ eV are represented. These peculiarities have shapes of
consequent distinct deeps of a widths $dE_h/E_h \simeq 0.2$ and of
relative amplitudes $dL/L \simeq {0.1 - 1.0}$, and are difficult
to be explained via known astrophysical mechanisms of particle
generation and acceleration.

  In the basis of the model lies the destruction of the Primary Cosmic
Radiation (PCR) nuclei on some monochromatic background of
interstellar space, consisting of the light particles of the mass
in the area of 36 eV ( maybe the component of a dark matter)
\cite{1} - \cite{4}. The destruction thresholds of PCR different
nuclear components correspond to the peculiarities of $E_h(E_0)$.
In this work the results of the recent treatment \cite{5} of large
statistical material are analyzed.  The experimental results are
in good agreement with the Monte-Carlo calculations carried out in
the frames of the proposed model.


\section{\bf Introduction}
\vspace{1pt}

  ~~~~In  \cite{1} - \cite{4} the possibility of the existence
of the Primary Cosmic Radiation (PCR) nuclei spectra cutoff effect
(i.e.  sharp decrease of their flux) in the energy range
$>\sim10^{16}$ eV was considered. The possibility to connect this
phenomenon with the destruction of the PCR nuclei on some diffuse
monochromatic background of the interstellar space consisting of
the particles of the mass  $\sim 30$ eV was discussed. In \cite{1}
the results were based on the investigation of high energy
multiple muons (so called muon groups) energy behavior in cosmic
ray Extensive Air Showers (EAS) as a function of $N_e$ (EAS size)
or of $E_0$  (primary energy of EAS initiating particle). In
\cite{2} - \cite{4} the energy $E_h$  of EAS hadronic component as
a function of $N_e$ or $E_0$ was investigated. The conclusions on
the existence of PCR nuclei spectra cutoff in the $N_e$  range
$10^{6} - 10^{7}$ particles (corresponding to the primary energy
range of the order of $10^{15} - 10^{16}$  eV) were confirmed.
{\footnote { We will not discuss here the other, in principle
possible but rather unnatural mechanisms to explain such a
structures in spectra.}} So the two different manifestations of
the same phenomenon were fixed by means of two different
installations using completely different methods.

    In the present work the results of the recent treatment of large archive
statistical material, carried out in \cite{5}, are analyzed. The
data were obtained by means of multilayer ionization calorimeter
(the area of each layer 36 $m^{2}$)  of the Tian- Shan high
mountain installation in the period before 1980, and were
retreated anew in the frames of the new approach. The work
\cite{5} was aimed at investigation of EAS hadronic component
energy using different statistical material, obtained at different
configurations of the installation, at different triggering
conditions and different selection criteria. Usage of an improved
approach to the data handling enabled us to establish the EAS
parameters with sufficiently high precision and to increase the
$E_h(N_e)/N_e$ dependence irregularities detection reliability and
the accuracy of their localization along the $N_e$ (or $E_0$)
axis.

  We will show that all these peculiarities may be naturally connected
with the main groups of PCR nuclei spectra consequent cutoff
process as a result of their destruction on some monochromatic
diffuse background of interstellar space and with the accumulation
of the products of this destruction.

  In section 2 we give the brief description of the main principles of
the analysis. In sections 3 and 4 the obtained experimental
results are discussed, the main parameters of the model and the
results of Monte-Carlo calculation of the observed effects made in
the frames of the proposed model are given.

\section{\bf{The experimental data and the basic principles
                of the analysis}}
~~~~As it was shown in \cite {2} -\cite{4} the experimentally
observed dependence of EAS hadronic component energy $E_h$ on EAS
size $N_e$ has noticeable peculiarities of behavior.

   The results of the last updated treatment of the large statistical
material (more than 350 000 handled events in total) are
reproduced from \cite{5} in  Fig. 1.

   On this Figure are represented the specific energies
$E_h(N_e)/N_e$ of the EAS hadronic component, obtained at the
different configurations of the experimental apparatus and at
different selection criteria. For details see \cite{5}.

   The lower borders by $N_e$ are determined by the material available
to-day and by triggering conditions (curves 1,4 and 5), or by
conditions of preliminary selection (curves 2 and 3). The upper
borders are determined by statistical restrictions due to the high
steepness of the EAS spectrum $I(N_e)$. The error bars allow for
statistical errors. The errors of  $N_e$   determination are of
the order of 10\%. The difference of mean slopes of the curves to
the left from the values $N_e \sim 10^{6}$ is easy to explain by
the difference in triggering conditions and in the selection
criteria. The solid curves (splain), are drawn through the
experimental points.

  On all the presented curves in the region of $N_e > 5\cdot10^{5}$
the series of deeps with decrease of $E_h$  are clearly seen. Most
clearly these irregularities are revealed in the region $N_e >
2\cdot10^{6}$ , where the identical localization of these
phenomena on all shown curves may be easily traced. The indicated
deeps in $E_h(N_e)/N_e$ dependence are located in the regions of
the same values of $N_e$ , regardless of experimental material
used and of triggering and selection conditions (curves 1,2 and
3). Moreover, the localization of these deeps does not depend on
the zenith angle of the event registration (curves 4 and 5 - the
angles $0^o - 20^o$ and $20^o - 30^o$ respectively).

\begin{figure}[h]
\centering
\begin{picture}(170,130)(100,240)
\includegraphics[scale=.6]{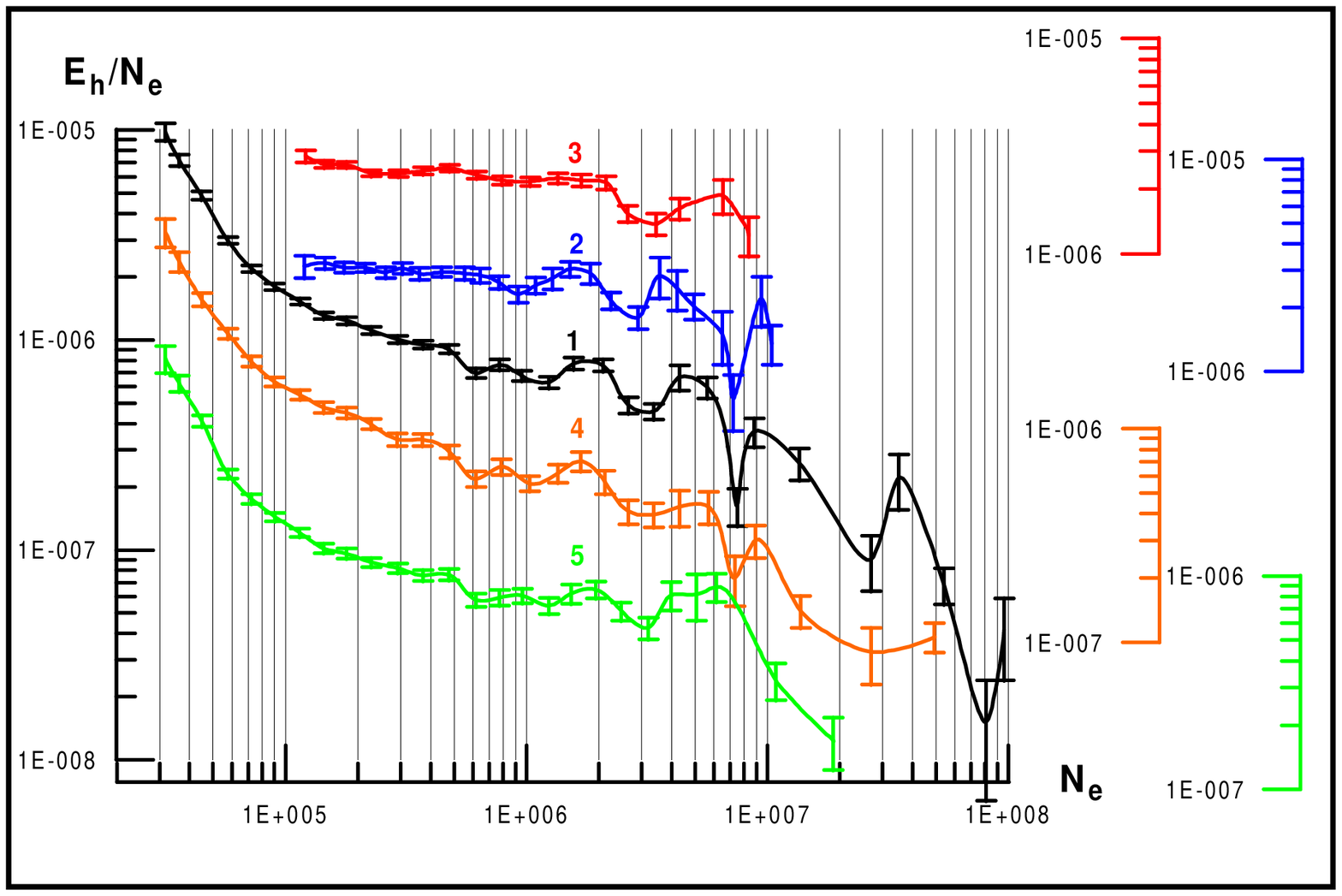}
\end{picture}
\parbox{10cm}{\vspace{10mm}{\bf Fig.1~:}~~ The specific energies
$E_h(N_e)/N_e$   of the EAS hadronic component for the different
configurations of apparatus and selection conditions. Details are
given in the text and in \cite{5}.  The left vertical scale
(arbitrary units) is given for curve 1. The other curves are
slightly shifted up or down to separate them from one another and
to clarify the picture. The corresponding parts of vertical scales
are shown at the right edge of the figure. }
\end{figure}

 In the region of relatively low values of $N_e$, namely at
$5\cdot10^{5} < N_e < 2\cdot10^{6}$ the mutually equivalent
irregularities may be seen as well. Here, however, a further
increase of accuracy and of statistics is required. In the region
$N_e  < 5\cdot10^{5}$ at this stage of investigation the
$E_h(N_e)/N_e$ dependence behavior is revealed as completely
smooth. The very last deep on the curve 1 is not provided
statistically.

  Actually, here we are dealing with the results obtained in the
several independent experiments.

  Apparently the sufficiently natural explanation for the creation
of such behavior of EAS hadronic component energy flux, proposed
by us in \cite{1},\cite{3} is based on the existence of PCR nuclei
spectra consequential cutoff in the primary energy region
$>10^{15}$  eV. The destruction of PCR nuclei is caused by their
interactions with some light particle (maybe the component of a
dark matter, since the neutrino does not match this role due to
its parameters) forming the monochromatic, low temperature
background in interstellar space.

 The results of the calculation of $E_h(N_e)/N_e$ dependence based
on this model are presented in Fig.3 and are discussed in section
4.

 According to this model, the cutoffs of PCR nuclei spectra have
a threshold character and depend on the primary energy of the
nucleus and on the energetic threshold of its destruction. For the
heavy nuclei of Fe group this phenomenon takes place at $N_e$ -s
higher then $\sim 7.4\cdot10^{6}$, which corresponds to primary
energies of the order of $1.4\cdot10^{16}$ eV. At the energies
higher then $\sim 4\cdot10^{16}$, at which the nuclei of Pb group
are destructed, the protons are dominant in the PCR, with the
exception of some part of survived nuclei and of the heaviest
nuclei like uranium.  The latter are destructed at the higher
energies.

  The essence of the deep formation mechanism on $E_h(N_e)/N_e$
dependence was given in \cite{5}. We reproduce it here with some
additional details.  The sketch illustrating this mechanism is
given in Fig.2.
\vspace{25mm}
\begin{figure}[h]
\centering
\begin{picture}(170,130)(50,0)
\includegraphics[scale=.60]{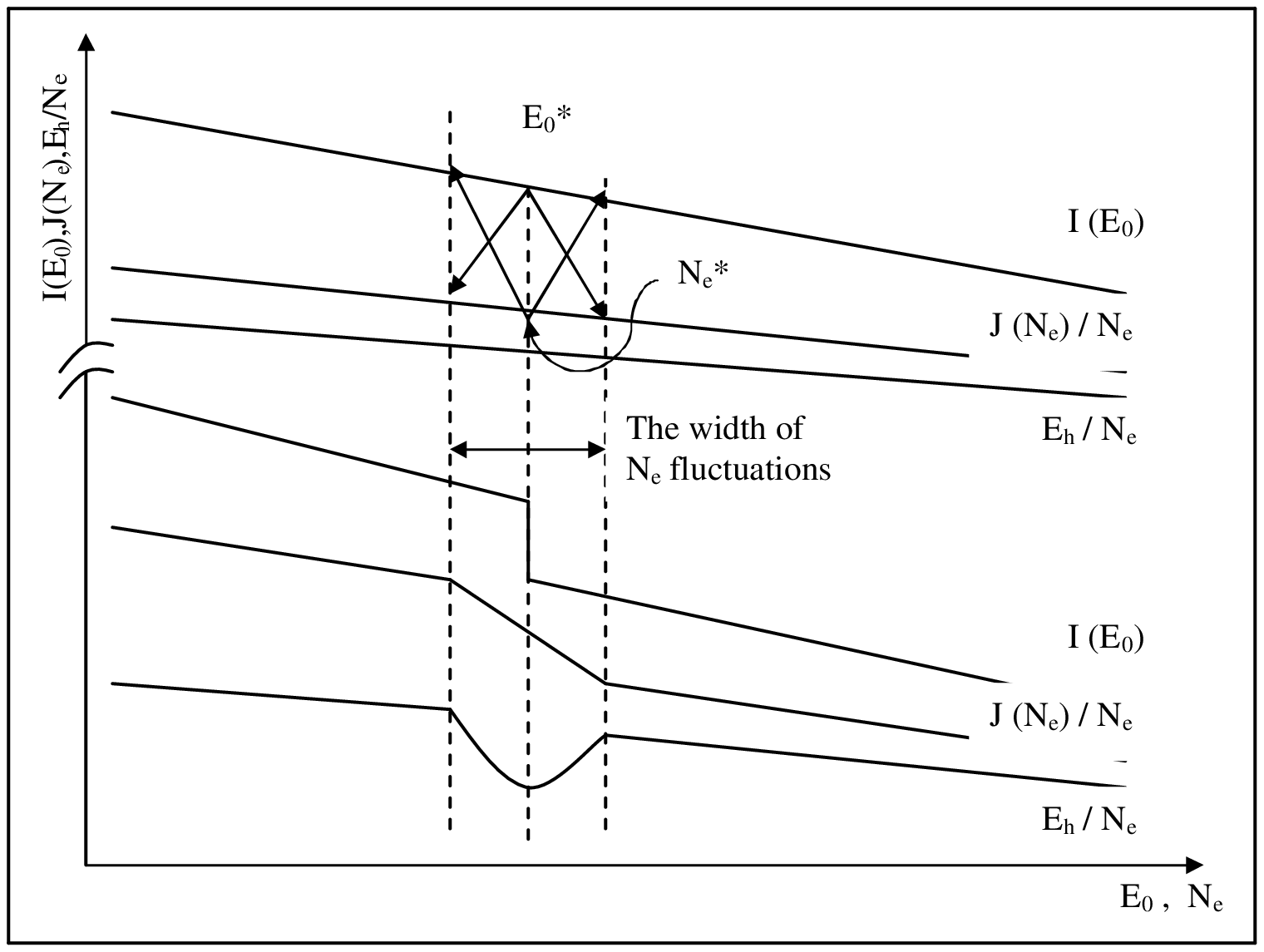}
\end{picture}
\parbox{10cm}{\vspace{10mm}{\bf Fig.2~:}~~ The sketch to explain
the mechanism of deep formation  }
\end{figure}

 The EAS of the certain given $N_e^\ast$ are generated by the primary
particles of the energy wedged within a rather wide interval of
primary energies $E_0$. There is some balance of the contributions
of these $E_0$, which depends on the slope of the primary energy
spectrum. On the other hand, at a fixed value $E_0^\ast$ of
primary particle energy, the number $N_e$ of particles in EAS
fluctuates within a rather wide range with the certain mean value
$N_e^\ast$.

  Let the  spectrum of the given certain component of PCR have the
form $I(E) = K \cdot E^{-r}$. Now let the spectrum of this
component of PCR be cut off above some fixed value of primary
energy $E_0^\ast$ (cutoff energy), i.e. at primary energies higher
then $E_0^\ast$ the flux of this component of PCR sharply
decreases (the value of coefficient $K$ falls sharply). However,
it is clear from the above that the spectrum of the corresponding
EAS will not be cut off above the value $N_e^\ast$ since the
showers from primary particles of the energies $E_0 < E_0^\ast$
will still be registered due to fluctuations of $N_e$ in the
showers of total energy $E_0 < E_0^\ast$. The showers initiated by
the primary particles (of the PCR component under consideration)
of the energy $E_0 > E_0^\ast$ will be present in extent of the
new value of coefficient $K$.

 It is established that the mean energy flux $E_h$ of hadrons
rises in average with the rise of $E_0$ (i.e with the rise of
$N_e$). Besides, at the fixed energy $E_0$ the anticorrelation of
$E_h$ and $N_e$ fits naturally with the account of the approximate
conservation of the sum of $e$- and $h$-component energies. It
follows here from, that in the case of spectrum cutoff the flux of
hadron energy in EAS of $N_e > N_e^\ast$ will decrease on average
as compared with the case of the absence of cutoff, since the EAS
for such $N_e$-s  now will be generated by the primary particles
of the energies $E_0 < E_0^\ast$. The balance of contributions in
showers of some given $N_e$ generated by primary particles of
different energies will be upset. The width of the interval in
which this effect does occur will be of the order of width of the
fluctuations of $N_e$ . At higher $N_e$-s exceeding the frames of
this interval, all manifestations of the flux of this component
will decrease sharply, and the spectrum of showers from this
component will fall as well. However, above the indicated interval
the balance of contributions is restored ( because the spectrum
has the same slope again) and the mean energy of hadrons in EAS
(not the intensity of EAS!) reaches its previous level. In the
case of complete cutoff of the spectrum EAS from this component
will vanish completely. For the mixed composition of PCR the
spectra of different components will be cut off at different
$E_i^\ast$ -s ( $i$ - indexes of the components) \cite{3} and one
has to expect deeps on the $E_h(N_e)/N_e$ plot near the
corresponding values of $N_e$ - s. The widths of these deeps are
determined by the ranges of $N_e$ fluctuations.

   In the region of $N_e < N_e^\ast$  the upset of the balance of
contributions to showers of given $N_e$  will take place as well.
Actually, in the absence of the cutoff effect the EAS of the
primary energy $E_0  > E_0^\ast$ are also contributing to this
region of $N_e$ . In case of the cutoff of the spectrum of a given
component this balance will be upset as well. The width of the
interval to the left of $N_e^\ast$  where this effect occurs is
also of the order of $N_e$ fluctuations. However, this effect will
be essentially less than the above considered one (due to the
steepness of the spectrum).

   This mechanism of   $E_h(N_e)/N_e$   dependence formation is
analyzed  in  \cite{3}, where the principles and results of the
calculation of the expected features of this dependence are given.
It is to be noticed however, that the level of data processing and
analysis in \cite{3} was lower then reached  to day, so the
numerical values of the model parameters and the localization of
the signals along the $N_e$ axis differ remarkably from the
present results.

     From the above presented qualitative description of the process, it
follows that the position of the deeps on the $E_h(N_e)/N_e$
dependence along the $N_e$ scale is determined by the PCR nuclei
threshold energies $E_0^\ast$, necessary to switch on the
mechanism of nuclei destruction on the particles of interstellar
monochromatic background.  The threshold energies of PCR nuclei
destruction are determined by kinematics of PCR nuclei destruction
in the interactions with the interstellar monochromatic background
particles of mass  $m$

\begin{equation}
A_0 + x =  A_1 + A_2 + x'
\label{1}
\end{equation}
where  $A_0$  is initial nucleus of PCR ,  $x$ - background
particle in the initial state,  $A_1$ and $A_2$ - nuclei-fragments
of destructed primary particle, $x'$ - background particle in
final state.

  For simplification we consider basically the 2-particle destruction
channels of nuclei, and besides they require the least amount of
energy transfer.

   The threshold energy $E = E_0^\ast$ for reaction(1) in the rest
system of $x$-particle is determined by the minimal value  of $S_0
= M_0^2  +  m^2  + 2Em$ , where $M_0$ and $E$ - are the mass and
the energy of the initial nucleus, consequently by the minimal
invariant value of $\omega^* = E \cdot m$, at which the reaction
(1) can take place.

  The invariant threshold  ~$\omega^\ast$ of the reaction (1) is
determined by the properties of the nucleus and its fragments - by
masses, binding energies etc. If $M_0$ - is the mass of the
initial nucleus, $M_1$ and $M_2$ - masses of nuclei- fragments,
while  $m$ - the mass of the background particle, then it is easy
to show that the reaction (1) can take place only under the
following condition

\begin{equation}
       E  > E^\ast = \omega^\ast/m = 1/2 \cdot [(M_1 + M_2 )^2 - M_0^2]
\label{2}
\end{equation}
which with good approximation may be written down in the more
convenient form

\begin{equation}
              E > E^\ast = \omega^\ast/m = (M_0 / m)\cdot\delta
\label{3}
\end{equation}
  where $\delta = M_1 + M_2  - M_0$ .

  So, at the fixed mass $m$ of the background particle these threshold cutoff
energies for starting of the destruction process of different
nuclei and for the different channels of the reaction (1)  will be
proportional to the corresponding invariant thresholds. Naturally
a sufficient degree of nuclei destruction is necessary to form
signals on the  $E_h (N_e)/N_e$ dependence - see section 3.

    To calculate the value of average hadron energy flux $E_h(N_e)$ in
the EAS of fixed particle number $N_e$ it is natural to use the
simultaneous possibility distribution of $E_h$ and $N_e$ in EAS at
the fixed initial energy $E_0$ . One can use any distribution
providing it corresponds with the EAS phenomenology established
for to day.
 Let   $P_i ( E_h, N_e | E_0)$  -  be the density of such distribution
for EAS of the initial energy $E_0$, where  $i$ = 1,2,3,......M  -
indicates one or another component of PCR, i.e. the kind of
nucleus under consideration. Then for the case of M kinds of PCR
the mean energy of the EAS hadronic component may be presented as
the average of specific mean energy fluxes. It may be calculated
under different assumptions on the behavior of PCR separate
components, in particular at the presence of the cutoff process of
these spectra at some threshold energies. The version of such
calculation using the lognormal distribution as  $P_i ( E_h, N_e |
E_0)$   is described  in \cite{3}. In the same work the
analytically calculated shape of $E_h(N_e)$ distribution is
presented, matched to the results obtained at that time.

 In this work we present the results of new calculations carried out by
means of the Monte-Carlo method (as one more corresponding to the
character of the task) based on the model proposed.

\section{\bf\bf{The analysis of nuclei destruction chains}}

 ~~~~Let us return now to Fig. 1, where the results of the work
\cite{5} are reproduced. They are represented in the form
$E_h(N_e) /N_e$ , i.e. the value of EAS hadronic component energy
flux per one electron of EAS.  As was said above, on all of the
given curves the deeps are clearly seen, located at the same
values of $N_e$. These deeps we bring further into correspondence
with the PCR nuclei destruction, considered in section 2.

  The consideration we'll carry out uses the curve 1 as one mostly
provided statistically and covering the widest diapason of the EAS
sizes $N_e$ (i.e. of $E_0$).

  At the first stage of consideration it will be natural to choose
the nucleus of a Fe as a bench-mark nucleus, the destruction area
of which $E_0 \sim 2\cdot 10^{16}$ eV was estimated in [1] most
reliably. The choice of Fe nucleus is determined also by the fact
that it is represented in PCR in a sufficiently large amount at
lower energies and so its destruction signal must distinctly stand
out against the background of the neighbor nuclei. And what is
very important, the fluctuations of $N_e$ in EAS initiated by Fe
nuclei are smaller than in those initiated by more light nuclei.
For these reasons the signal of Fe destruction must be less
diffuse as compared with the signals of the other nuclei or their
groups. On the strength of the aggregate of these indications the
deep with the minimum at $N_e = 7.4 \cdot 10^6$ was taken as a
signal of Fe destruction (the corresponding value of $E_0$ is
$\sim1.4\cdot10^{16}$ eV). It may be distinctly seen on the curves
1, 2 and 4. On the curves 3 and 5 these ranges of $N_e$ - s are
not provided with the statistical material. It is natural to take
for the $N_e$, corresponding to the threshold energy of Fe
destruction, the value $N_e = 7.4\cdot10^{6}$ where the signal
under consideration has a minimum. The location of $N_e$ threshold
value, in the minimum point of the signal, obviously follows from
the above given qualitative description of the process.

  Taking the indicated point as a bench-mark, and with account of the
value of lowest invariant threshold of Fe destruction, one can
evaluate the mass of the background particle on which the
destruction of Fe nuclei takes place.

  Here however one must take into account the necessity of the
achievement of a sufficient degree of nuclei destruction, to let
the heaviest fragments be shifted far enough to the left from the
threshold energy of the initial nucleus. The value of this shift
must provide the exit of the signal area out of the limits of
possible $N_e$ fluctuations in EAS, caused by the most massive
fragments at the end of the destruction chain. This is necessary
to exclude all traces of the influence of destructed nuclei and
their fragments on $E_h(N_e)/N_e$ dependence formation in the
region of the signal. Without this, the above described mechanism
of signal creation on $E_h(N_e)/N_e$ dependence will work less
efficiently. Assuming  maximum possible fluctuations have the
value $\sim50\%$ (including the measurement error) one can
conclude that the mass of the heaviest fragment must not exceed
the value of about 0.5 of the mass of the initial nucleus. In the
case of Fe this corresponds to the nucleus of Si28.

  Now it is easy to extract the value of the background particle mass
$m = 36.76$ eV (in [3] we represented the value of $m\sim36.4$ eV)
which provides the existence of the whole chain of destructions
from Fe56 down to Si28, started from the above indicated
experimental value of $N_e = 7.4\cdot10^{6}$ for  the Fe56
destruction signal ($E_0 = 1.42\cdot10^{16}$ eV). In reality,
under these conditions the much deeper continuation of the
destructions chain is automatically provided. Note, that by
approaching the indicated value of $E_0$, more and more deep
consequent links of the destruction chain are joining to it,
resulting in noticeable flattening of the left fronts of the
signals, which can be clearly seen on the experimental curves.The
steepness of the right front is noticeable higher. It is
determined by the steepness of the spectrum cutoff and by the
accuracy of measurements, of course.

 After this, proceeding from the obtained value of $m$  and from the
invariant destruction thresholds of the other nuclei, it is easy
to distribute the areas of their destruction along the $N_e$ scale
(providing corresponding chains of destructions). The result is
shown on Fig. 3. The vertically downwards directed arrows
symbolize the destruction regions of main stable nuclei of PCR
from He4 to Fe56. It must be noted that such a correspondence
between the PCR and signals on the $E_h(N_e) /N_e$  dependence
turn out to be the best one. In any other version the level of
correspondence becomes explicitly unacceptable.

  The numeric values of the nuclei destruction thresholds, as well as
the threshold values of $N_e$ and $E_0$ for the basic nuclei of
PCR, are quoted in~Table 1. The data for some of the light nuclei
are quoted as well.

Note that for all nuclei quoted in Table 1, the energetic
thresholds for the destruction down to fragments of the mass $M
<\sim 0.5\cdot M_0$ provide the destruction down to He4 as well.
The values of the masses for the nuclei $A_0$, $A_1$ and $A_2$,
entering the expression (1) are calculated in accordance with the
Tables of Atomic Masses \cite{6}. For the heavy nuclei of the Pb
the calculation was made approximately. The invariant threshold of
N14 destruction, for which the binding energy per nucleon (7.7
MeV) is close to corresponding value for Pb (7.9 MeV), was assumed
as a basis and then the correction was allowed proportional to the
relation of the mean binding energy and of the masses of these
nuclei.

\newpage
\begin{center}
\begin{picture}(570,550)(100,120)
\includegraphics[scale=.95]{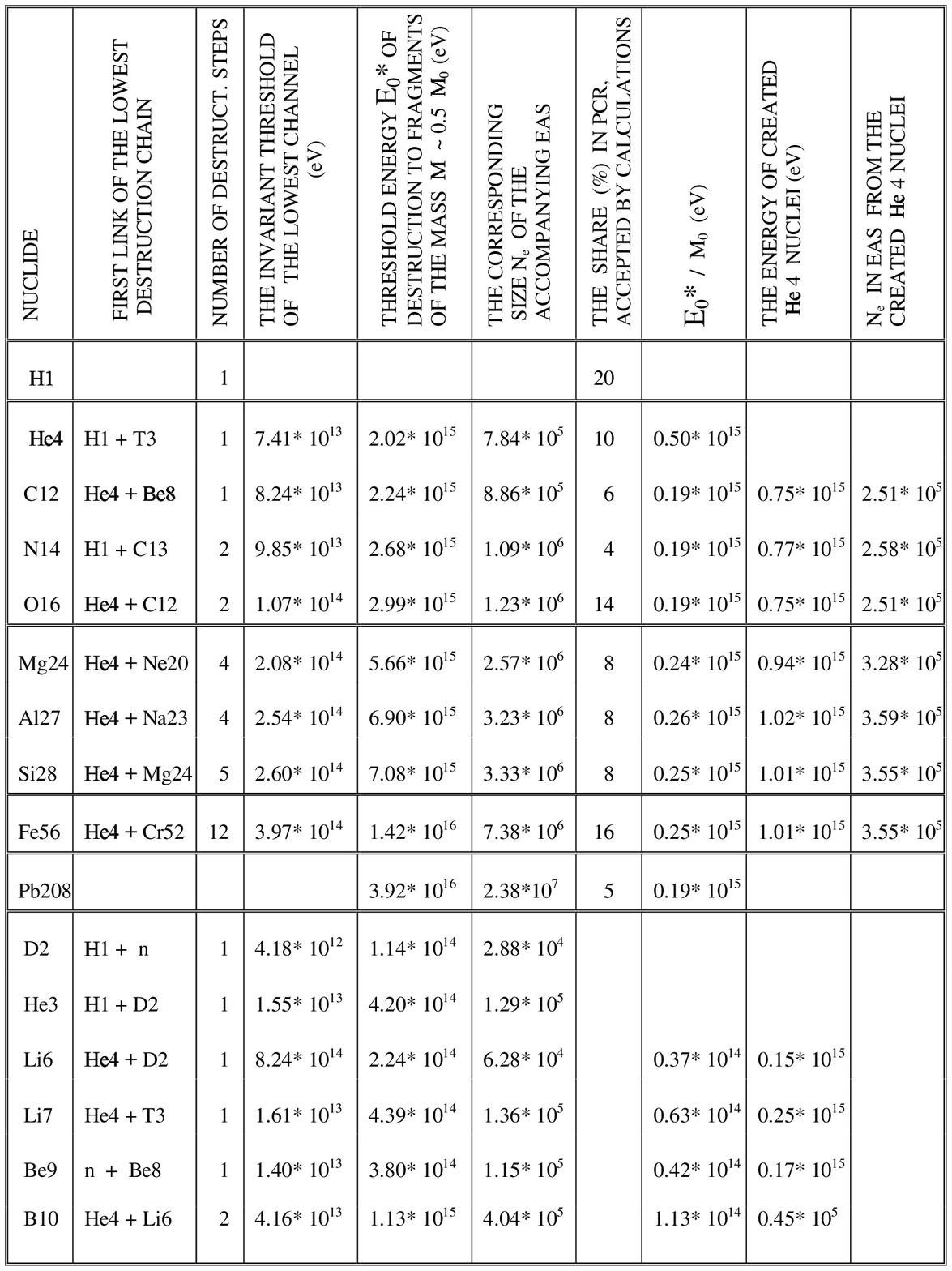}
\end{picture}
{\bf \Large Table 1 }
\end{center}

\newpage

  From Table 1 it can easily be seen that for the majority of nuclei
the threshold values of $N_e$ may be divided into groups disposed
on the graph near the areas of the main peculiarities of the
$E_h(N_e)/N_e$ dependence. Division into such groups can be easily
explained by the influence of the periodicity (oscillations) in
the dependence of alpha-particles binding energy in the nucleus on
its mass. The groups are formed by nuclei with the close values of
the quantity  $E^\ast = \omega^\ast / m$.
  The obtained value of the background particle mass can be expressed
as \\
$$
                         m ~=~  36.76 \pm (10 + 5) \%  ~~eV
$$
    Here the value in parenthesis is the sum of systematic error ,
determined by the accuracy of the transition from the experimental
value of the particle number  $N_e$ in EAS to the energy $E_0$ of
the initial nucleus  (evidently it is of the order of 10 \%) and
of the statistical error, not exceeding  $\sim 5$ \% .

 In articles \cite{1} and \cite{3} we presented the value of $m$ of
the order of 30 eV and 36.4 eV . The refinement of the signal
localization along the $N_e$ axis and usage of the new approach
led to a close value of $m$.

 It must be noted however, that there are not any peculiarities
corresponding to the destruction of nuclei of B10, Be9, Li6, Li7
and most light nuclei of He3 and D. It can be the result of a too
small amount of these nuclei in PCR, which is below the resolution
level limit achievable at the recent stage of data treatment. It
would be extremely interesting and important to fix the signals of
He3  in the $N_e$ region  $\sim1.3\cdot10^5$ , B10 in the $N_e$
region $\sim 4\cdot10^5$ and especially  D2 in the $N_e$ region
$\sim 2.9\cdot10^4$, which must be located on the graph completely
apart from the other nuclei. The discovery at least of some of
these signals would confirm the above described effects with very
high reliability. The efforts to increase the precision level of
the information extraction are now in progress and seem to have
realistic prospects.

 \section{\bf \bf{Monte-Carlo calculation of the disturbed flux of
hadron energy}}

 ~~~~In Fig.3 the result of the Monte-Carlo calculation of
$E_h(N_e)/N_e$ dependence, carried out in the frames of the
proposed model, is shown. The nuclei from Table 1, as  those most
abundant in PCR, are included in the calculation.
%
\begin{figure}[h]    \centering
\begin{picture}(130,125)(50,0)
\includegraphics*[scale=.55]{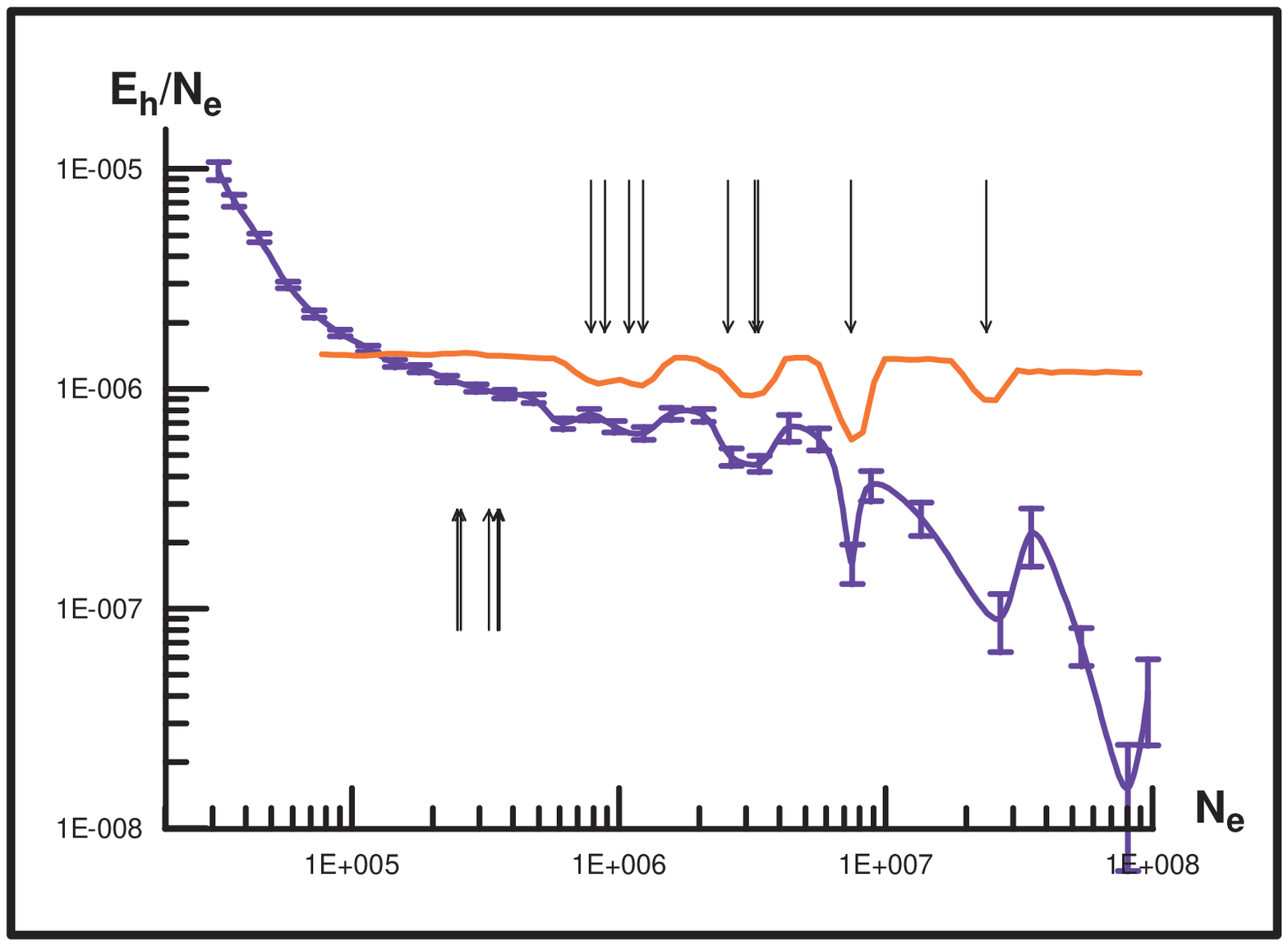}
\end{picture}
\parbox{10cm}{\vspace{10mm} {\bf Fig.3~:} ~~ The blue curve and the
downward directed arrows - the correspondence between the deeps on
the dependence $E_h(N_e)/N_e$ and the thresholds of PCR nuclei
destruction. The upward directed arrows - the low energy border of
the area of PCR nuclei fragments accumulation. The red curve - the
Monte-Carlo calculation of the dependence $E_h(N_e)/N_e$ in the
frames of the proposed model. }
\end{figure}
The shares of the nuclear components in PCR, accepted by the
calculations, are also given in Table 1. For the fluctuations of
$N_e$ and $E_h$ in EAS, initiated by these nuclei, the values
respectively 25\% and 100\% are accepted. Note that in the value
of $E_h$ fluctuations not only the necessity of energy balance
conservation between the EAS components contributes, but the
variation of the mean longitudinal momentum as well, which results
in the variation of the secondary hadrons front cone and
consequently in the variation of the density of hadron flux
through calorimeter. As a result the hadron energy flux through
the calorimeter fluctuates more sharply. It is obvious from Fig.3,
that the proposed model easily and naturally leads to the observed
effect.

    As can be seen from Table 1, to destruct any of the quoted
nuclei only about one collision with the transfer of required
energy per each 4 nucleons of the nucleus is needed. So as a
condition of the process under consideration one can write
\begin{equation}
          \sigma= n\cdot(L\cdot\rho)^{-1} cm^{-2} ,
\label{4}
\end{equation}

where $n=0.25$ is the number of the effective collisions per
nucleon of the nucleus, $L$ is the length of the nucleus
propagation in space,  $\rho$ is the density of background
particles in space, $\sigma$ is the cross-section of nucleon
interaction with the background particle.

    This order of cross-section seems to be the optimal one, because at
$n = 1$ all nucleons (including the singles) will interact at
least once, which would lead to the  PCR spectrum  cut off already
at the energies of the order of $10^{17}$ eV due to the energy
losses of nucleons caused by multiple production of pions on
background particles. (The process analogous to GZK process, which
takes place at the energies  $>\sim 5\cdot10^{19} eV)$. In case of
the cross-section written above only of $\sim 25\%$  of protons
will be involved in the process. This process, together with the
nuclei destruction and accumulation of their fragments at lower
energies can lead to the experimentally observed steepening of
primary spectrum at the energies $>\sim 4\cdot10^{15} eV$.

  In this connection one has to note the following circumstance as
well. The majority of the lowest invariant channels are those with
the separation of the alpha - particle. It is important to notice
as well, that at the described process of nuclei destruction the
specific energy per one nucleon of the nucleus is practically
preserved. As can be seen from Table 1, the specific threshold
energy of He4 destruction is about 2 - 3  times higher than that
of the other nuclei. From here it is easy to see that He4,
generated as a result of the destruction of heavier nuclei at the
energies not exceeding the corresponding thresholds more than
about 2 – 3  times, will not be destructed at the consequent
impacts with the background particles.  The whole of this He4 will
be accumulated in the area of lower energies, determined by the
relation between the masses of He4 and the destructed nucleus (see
the last column of Table 1).  This low energy border of the
fragment accumulation area of all the nuclei participating in the
process is shown in Fig.3 by upward directed arrows.  However, due
to high steepness of the energy spectrum and dispersion of
fragments by the energy we have no effects of threshold character
here, and the expected influence on the shape of $E_h(N_e)/N_e$
dependence is weak. Nevertheless, despite the smooth character of
fragment accumulation, their amount in the accumulation area can
reach up to  $>\sim 5\%$ of the primary spectrum according to the
preliminary evaluation, and one can expect the influence of this
process on the shape of the spectrum. It is not excluded that
namely the aggregate of all these processes, including the
accumulation of the fragments of destructed nuclei, leads to the
macro effect - the well known peculiarities of the primary
spectrum in the energy region $10^{15}$ - $10^{17}$ eV - the so
called "bump" on the primary spectrum in the area $\sim10^{15}$ eV
and the change of PCR spectrum slope in the consequent region.
Preliminary calculation leads to qualitative accordance with the
observed shape of the primary spectrum.

 \section{\bf Conclusions}

 ~~~~In conclusion let us formulate the main obtained results.

$\bullet$  The experimentally observed irregularities of the EAS
muon (muon groups) \cite{1} and hadron \cite{2} - \cite {5}
components energy dependence on the energy of the initial
particle, can be naturally treated as a result of  PCR nuclei
spectra cutoff in the primary energy region $10^{15} - 10^{17}$
eV.

$\bullet$   These irregularities are difficult to explain by means
of the "ordinary" astrophysical mechanisms, because the
contributions from the different sources of cosmic rays will "slur
over" all small irregularities of the spectrum.

$\bullet$   From our point of view, the most attractive model is
one where this effect is a result of nuclei destruction caused by
their interaction with some monochromatic background of
interstellar space, consisting of the particles of the mass $m$ in
the area of 36 eV.

$\bullet$  The Monte-Carlo calculation of the $E_h(N_e) / N_e$
dependence based on the proposed model gives  satisfactory
accordance with the experiment. The calculation was performed
without account of the particular properties of the apparatus. To
simplify the calculations we admitted, that the mechanism of
nuclei destruction on interstellar monochromatic background is
determined by threshold expressions  $E^\ast = \omega^\ast/m$ ,
where $\omega^\ast$ is the invariant threshold of initial nuclei
destruction by collisions with the background particle of the mass
$m$. At the same time the dependence of process cross-sections on
the energy was ignored. In other words we considered the distance
$L$ from sources to Earth much larger than the mean path of nuclei
for interaction with background particles "m". We considered the
interaction with "m" - particles as a local one, and not differing
between neutrons and protons, and so kinematically similar to low
energy lepton-nuclear interactions.

     So the minimal condition of the cross-section of this process
is:\\

                 $\sigma = n\cdot(L\cdot\rho)^{-1} cm^{-2}$  ,\\   \\
where $n=0.25$ is the number of the effective collisions per
nucleon of the nucleus,   $L$ is the length of the nucleus
propagation in space, $\rho$ is the density of background
particles in space, $\sigma$ – is the cross- section of nucleon
interaction with background particles.

 Let us evaluate the order of value $\sigma$. For $m =\sim 30$ eV
the maximum admissible   $\rho$ = 30 - 40 particles per $cm^3$ –
from the restrictions on the density of the dark matter, not
exceeding the value $1-1.5$ KeV per $cm^3$. The maximum realistic
$L = 3\cdot10^{17} sec\cdot3\cdot10^{10} cm/sec = 10^{28} cm$
corresponds to 0.1 of the Universe age. So we have $\sigma >\sim
n\cdot(10^{28}\cdot\rho)^{-1} =\sim 10^{-3}$ mb. If one supposes
the possibility of the accumulation of these particles in the
Galaxy, then one can increase in principle the density $\rho$ not
more than by 3 orders (starting from the mean size of galaxies and
from the mean distances between them). This however will lead to
more rigid restrictions on the radiation life time, at least by 1
order. This means, that even at these conditions the value of
$\sigma$  will be $\sim 10^{-4}$ - $10^{-5}$ mb. This is a
sufficiently large value, so the processes with the participation
of "m" - particles would be observed in the acceleration
experiments with high probability. However, to date this has not
taken place. If the above discussed interpretation of experimental
results is correct then the "m" - particle has to be a
neutrino-like particle. However the massive one ( of the type of
the 4-th generation of neutrino ), participating only in the
processes with neutral currents.

$\bullet$ The macro effect probably caused by the aggregate of the
described processes   - nuclear spectra cutoff and shift of the
particles by energy - is the so called "bump" on the primary
spectrum in the energy region of the order of $3\cdot10^{15}$ eV
and the steepening of the spectrum in the consequent diapason of
the energies.

$\bullet$ If the approach proposed in this work is correct, then
the further increase of the statistics and of the data handling
precision will possibly allow us to try to fix the destruction
signals of B10, Be9, He3 and especially of D2. The weak signals of
Li6 and Li7 may be possibly observed from the channel of their
destruction down to He4 (the mass of He4 accounts for more than
50\% of the mass of Li6 and Li7, and because of that the signals
of their destruction will not be so clear). The threshold energies
of the above listed nuclei destruction differ sharply from those
for other nuclei, and because of that the signals of their
destruction are located apart on the axis of primary energies (or
$N_e$ axis). The discovery at least of some of these signals would
give us the additional bench mark which will allow estimating the
value of the background particle mass with higher reliability and
precision. The expected places of these signals on the $N_e$ scale
(with the accuracy of $\sim {15} \%$ , determined by the accuracy
of $N_e \longleftrightarrow E_o$ transition) are
$N_e\sim1.3\cdot10^5$ ~for He3 , $N_e\sim1.15\cdot10^5$ ~for Be9,
$N_e\sim4\cdot10^5$ ~for B10 and $N_e\sim2.9\cdot10^4$ ~for D. The
expected area of the location of the comparatively weak signals of
Li6 and Li7 are $N_e \sim 6.3\cdot10^4$ and $N_e \sim
1.4\cdot10^5$ respectively.

 The above given evaluation of the cross-section of the interaction
with background particle makes it in principle possible for
generation of the signals from the next energetic channels of
destruction. However for the statistics available to date, the
probability of extracting these signals or of observing the "fine
structure" of the signals from the groups of nuclei i.e. to
extract the signals of the separate nuclei inside the groups (or
subgroups of the groups) of nuclei is small, though some
indication of the existence of such a structure can be seen on the
signal of  C,N,O group in the region of $N_e \sim 6 \cdot 10^5
\div 10^6$
(curve 1 in fig 1).\\

\newpage

\noindent{\bf Acknowledgments}\\

 ~~~~The authors express their deep gratitude to O.V.Kancheli for his
interest in this investigation, numerous fruitful discussions and
advice. The authors are sincerely grateful  to Yu.D.Kotov for
discussion, J.M.Henderson - for  interesting discussions and his
assistance in preparation of this paper.

  This investigation is in part supported by a grant from the
Georgian Academy of Sciences and by the firm Mantex computers ltd.
(Tbilisi).


\end{document}